\documentclass[aps,prl,showpacs]{revtex4}
\usepackage{amssymb}
\usepackage{amsmath}

\usepackage{graphicx}
\usepackage{bm}

\begin{document}

\title{Quasi-exact-solution of the Generalized $E\otimes \varepsilon $
Jahn-Teller Hamiltonian}
\date{\today}
\author{Ramazan Ko\c{c}}
\email{koc@gantep.edu.tr}
\affiliation{Department of Physics, Faculty of Engineering University of Gaziantep, 27310
Gaziantep, Turkey}
\author{Hayriye T\"{u}t\"{u}nc\"{u}ler}
\email{tutunculer@gantep.edu.tr}
\affiliation{Department of Physics, Faculty of Engineering University of Gaziantep, 27310
Gaziantep, Turkey}
\author{Mehmet Koca}
\email{kocam@squ.edu.om}
\affiliation{Department of Physics, College of Science, Sultan Qaboos University, PO Box
36, Al-Khod 123, Muscat, Sultanate of Oman}
\author{Eser K\"{o}rc\"{u}k}
\email{korcuk@gantep.edu.tr}
\affiliation{Department of Physics, Faculty of Engineering University of Gaziantep, 27310
Gaziantep, Turkey}

\begin{abstract}
We consider the solution of a generalized $E\otimes \varepsilon $ Jahn-Teller
Hamiltonian in the context of quasi-exactly solvable spectral problems.
This Hamiltonian is expressed in terms of the generators of the $osp(2,2)$ Lie algebra.
Analytical expressions are obtained for eigenstates and eigenvalues. The
solutions lead to a number of earlier results discussed in the literature.
However, our approach renders a new understanding of ``exact isolated''
solutions.
\end{abstract}

\maketitle

\section{Introduction}

The Jahn-Teller (JT) distortion problem is an old one, dating back over
sixty years \cite{1}. In 1958 Longuet Higgins et al. \cite{2} carried out
the first diagonalization of the Hamiltonian of the $E\otimes \varepsilon $
JT system within the adiabatic approximation. Yet, even today, new
contributions to this problem are being made \cite{3}. They appear, however,
not to have been fully exploited in the analysis of the JT problem. The $%
E\otimes \varepsilon $ JT problem is a system with a doubly degenerate
electronic state and doubly degenerate JT active vibrational state. The JT
effect describes the interaction of degenerate electronic states through
non-totally symmetric, usually non-degenerate nuclear modes. This effect
plays an important role in explaining the structure and dynamics of solids
and molecules in degenerate electronic states.

Studies of the JT effect led Judd to discover a class of exact isolated
solutions of the model \cite{4}. To determine the relations between the
parameters of the model, one can obtain an analytic form of two eigenvectors
of the Hamiltonian corresponding to the specific energy. The complete
description of these solutions has been given by Reik et al\cite{5}. They
observed that isolated solutions can be obtained by using Neumann series of
expansions of the eigenvectors in the Bargmann-Fock space described by the
boson operators. The same problem has been treated by Klenner in polar
coordinates\cite{6}. The possibility of the existence of some exact
eigenvalues was previously proposed by Thorson and Moffitt \cite{7}. Their
results have been generalized by Loorits \cite{8}.

As a related topic, the concept of quasi-exactly solvable (QES) systems
discovered \cite{9,10} in the 1980s, has received much attention in recent
years, both from the viewpoint of physical applications and their inner
mathematical beauty. It turns out that in quantum mechanics there exist such
some systems; a part of their spectrum can be computed using algebraic
methods. The classification of the $2\times 2$ matrix differential equations
in one real variable possessing a polynomial solution has been carried out %
\cite{11}. The relevant algebraic structure is the graded algebra $osp(2,2)$%
. The $E\otimes \varepsilon $ JT Hamiltonian can be expressed as two coupled
differential equations. The $2$-dimensional QES matrix Schr\"{o}dinger
equation, with one real and one Grassmann variable have also been
investigated\cite{12,15}. More recently, we have shown that the $E\otimes
\beta $ JT Hamiltonian is quasi-exactly solvable \cite{16}. In this paper,
we present a quasi-exact solution of the $E\otimes \varepsilon $ JT
Hamiltonian.

This paper is organized as follows. In \S 2, following Reik's approach, we
express the $E\otimes \varepsilon $ Hamiltonian in the Bargman--Fock space
to obtain a second-order differential equation. In \S 3, this differential
equation is expressed as linear and bilinear combinations of the generators
of $osp(2,2)$ algebra. The condition for quasi-exact solvability of the $%
E\otimes \varepsilon $ JT Hamiltonian is then derived. In \S 4, we discuss
and compare our results with those studied given in the literature in
different contexts.

\section{The $E\otimes \protect\varepsilon $ Jahn-Teller Hamiltonian}

The well-known form of the $E\otimes \varepsilon $ JT Hamiltonian describing
a two-level fermionic subsystem coupled to two boson modes has obtained by
Reik\cite{5} is given by%
\begin{equation}
H=a_{1}^{+}a_{1}+a_{2}^{+}a_{2}+1+(\frac{1}{2}+2\mu )\sigma ^{0}+2\kappa
\lbrack (a_{1}+a_{2}^{+})\sigma ^{+}+(a_{1}^{+}+a_{2})\sigma ^{-}],
\label{eq:1}
\end{equation}%
where $\frac{1}{2}+2\mu $ is the level separation, and $\kappa $ is the
coupling strength. The Pauli matrices $\sigma ^{\mp ,0}$are given by%
\begin{equation}
\sigma ^{+}=\left[ 
\begin{array}{cc}
0 & 1 \\ 
0 & 0%
\end{array}%
\right] ,\quad \sigma ^{-}{}=\left[ 
\begin{array}{cc}
0 & 0 \\ 
1 & 0%
\end{array}%
\right] ,\;\sigma ^{0}=\left[ 
\begin{array}{cc}
1 & 0 \\ 
0 & -1%
\end{array}%
\right] ,  \label{eq:2}
\end{equation}%
and they satisfy the commutation relations%
\begin{equation}
\lbrack \sigma ^{+},\sigma ^{-}{}]=\sigma ^{0},\quad \lbrack \sigma
^{0},\sigma ^{\mp }]=\mp 2\sigma ^{\mp }.  \label{eq:3}
\end{equation}

The annihilation and creation operators, $a_{i}\;$and$\;a_{i}^{+},$ satisfy
the usual commutation relations,%
\begin{equation}
\lbrack a_{i}^{+},a_{j}^{+}]=[a_{i},a_{j}]=0,\quad \lbrack
a_{i},a_{j}^{+}]=\delta _{ij}.  \label{eq:4}
\end{equation}%
Following the analysis of Reik\cite{5}, we can express the Hamiltonian in
the form%
\begin{equation}
\frac{1}{2}H=\frac{1}{2}J+\frac{1}{2}+h,  \label{eq:5}
\end{equation}%
where $J$ represents the angular momentum of the system and is given by%
\begin{equation}
J=a_{1}^{+}a_{1}-a_{2}^{+}a_{2}+\frac{1}{2}\sigma _{0}.  \label{eq:6}
\end{equation}%
Note that $J$ commutes with $h,$ and the eigenvalue problem of the angular
momentum part can be easily solved. It reads%
\begin{equation}
J\left| \psi \right\rangle _{j+\frac{1}{2}}=\left( j+\frac{1}{2}\right)
\left| \psi \right\rangle _{j+\frac{1}{2}},\quad \left( j=0,1,2\cdots
\right)   \label{eq:7}
\end{equation}%
with the eigenfunctions%
\begin{equation}
\left| \psi \right\rangle _{j+\frac{1}{2}}=(a_{1}^{+})^{j}\phi
_{1}(a_{1}^{+}a_{2}^{+})\left| 0\right\rangle \left| \uparrow \right\rangle
+(a_{1}^{+})^{j+1}\phi _{2}(a_{1}^{+}a_{2}^{+})\left| 0\right\rangle \left|
\downarrow \right\rangle ,  \label{eq:8}
\end{equation}%
where $\left| 0\right\rangle $ is the vacuum state for both bosons. Here $%
\left| \uparrow \right\rangle $ and $\left| \downarrow \right\rangle $ are
the eigenstates of the operators $\sigma ^{0}$ , and $\phi _{1}$ and $\phi
_{2}$ are arbitrary functions of $a_{1}^{+}a_{2}^{+}$ . Because the
operators $h$ and $J$ commute, the eigenfunctions (\ref{eq:8}) are also the
eigenfunctions of the Hamiltonian (\ref{eq:1}). Therefore, we can write the
eigenvalue equation%
\begin{equation}
H\left| \psi \right\rangle _{j+\frac{1}{2}}=E\left| \psi \right\rangle _{j+%
\frac{1}{2}}  \label{eq:9}
\end{equation}%
and the equivalent Schr\"{o}dinger equation%
\begin{equation}
h\left| \psi \right\rangle _{j+\frac{1}{2}}=\epsilon \left| \psi
\right\rangle _{j+\frac{1}{2}},  \label{eq:10}
\end{equation}%
whose eigenvalue $\epsilon $ and $E$ are related by%
\begin{equation}
E=2\epsilon +j+\frac{3}{2}.  \label{eq:11}
\end{equation}

From the above consideration, it is found that the solution of the Schr\"{o}%
dinger equation (\ref{eq:1}) can be reduced to the solution of (\ref{eq:10}%
). The Hamiltonian $h$ can be expressed in the Bargman-Fock space by using
the realizations of the bosonic operators%
\begin{equation}
a_{i}^{+}=z_{i},\quad a_{i}=\frac{d}{dz_{i}},\;i=1,2.  \label{eq:12}
\end{equation}%
In this formulation, the Hamiltonian $h$ consists of two independent sets of
first-order linear differential equations. Substituting (\ref{eq:8}) and (%
\ref{eq:12}) into (\ref{eq:10}) and defining $\xi =z_{1}.z_{2}$ we can
obtain the following two linear differential equations satisfied by the
functions $\phi _{1}$ and $\phi _{2}$: 
\begin{subequations}
\begin{eqnarray}
\lbrack \xi \frac{d}{d\xi }-(\epsilon -\mu )]\phi _{1}+\kappa \lbrack \xi 
\frac{d}{d\xi }+(\xi +j+1)]\phi _{2} &=&0,  \label{eq:13a} \\
\kappa \lbrack \frac{d}{d\xi }+1]\phi _{1}+[\xi \frac{d}{d\xi }-(\epsilon
+\mu )]\phi _{2} &=&0.  \label{eq:13b}
\end{eqnarray}

These coupled differential equations represent the Schr\"{o}dinger equation
of the $E\otimes \varepsilon $ JT system in Bargmann's Hilbert space. In
what follows, we introduce a new variable, 
\end{subequations}
\begin{equation}
\xi =\kappa ^{2}(1+x),  \label{eq:14}
\end{equation}%
and redefine the functions%
\begin{equation}
\phi _{2}(\xi )=-\frac{1}{\kappa }\varphi _{2}(x),\quad \phi _{1}(\xi
)=\varphi _{1}(x)+\varphi _{2}(x),  \label{eq:15}
\end{equation}%
to obtain the following systems of differential equations: 
\begin{subequations}
\begin{align}
& [-x\frac{d}{dx}+(\epsilon +\kappa ^{2}-\mu )\varphi _{1}(x)+[-x\frac{d}{dx}%
+2\epsilon +2\kappa ^{2}+1+j+\kappa ^{2}x]\varphi _{2}(x)=0,  \label{eq:16a}
\\
& [\frac{d}{dx}+\kappa ^{2}]\varphi _{1}(x)+[-x\frac{d}{dx}+(\epsilon
+\kappa ^{2}+\mu ]\varphi _{2}(x)=0.  \label{eq:16b}
\end{align}%
It will be shown in the next section these equations can be solved in the
framework of quasi-exact solvability.

\section{QES of the Hamiltonian}

The general procedure to solve a differential equation quasi-exactly is to
express it in terms of the generators of a given Lie algebra having a finite
dimensional invariant subspace and use algebraic operations. In a recent
paper, \cite{16} we employed the $sl_{2}(R)$ Lie algebra in application to
the $E\otimes \beta $ system. In the present problem, the corresponding Lie
algebra is a subalgebra of the $osp(2,2)$ Lie algebra. Let us introduce the
following operators of the $osp(2,2)$ algebra\cite{15}: 
\end{subequations}
\begin{subequations}
\begin{eqnarray}
J_{+} &=&x,\quad J_{-}=x\frac{d^{2}}{dx^{2}}-2k\frac{d}{dx}+\sigma
^{-}\sigma ^{+},\quad   \notag \\
J_{0} &=&x\frac{d}{dx}-k+\frac{1}{2}\sigma ^{-}\sigma ^{+},\quad J=k+\frac{1%
}{2}\sigma ^{-}\sigma ^{+},  \label{eq:17a} \\
Q_{1} &=&\sigma ^{-},\quad Q_{2}=\sigma ^{-}\frac{d}{dx},\quad \bar{Q}%
_{1}=-\sigma ^{+}x\frac{d}{dx}+2k\sigma ^{+},\quad \bar{Q}_{2}=x\sigma ^{+}.
\label{eq:17b}
\end{eqnarray}%
Here $J_{i}$ ($i=0,+,-$) are the bosonic generators and $Q_{1,2}$ and$\;\bar{%
Q}_{1,2}$ are the fermionic generators. These generators form the $osp(2,2)$
algebra. The bosonic generators form $sl_{2}(R)$ algebra, 
\end{subequations}
\begin{equation}
\lbrack J_{+},J_{-}]=-2J_{0},\quad \lbrack J_{0},J_{\pm }]=\pm J_{\pm },
\label{eq:17c}
\end{equation}%
and the fermionic generastors satisfy the anticommutation relations 
\begin{equation}
\left\{ Q_{i},Q_{j}\right\} =\left\{ \bar{Q}_{i},\bar{Q}_{j}\right\}
=0,\quad i=1,\;2.  \label{eq:18}
\end{equation}%
The number operators of the system are given by%
\begin{equation}
N_{1}=\left\{ Q_{1},\bar{Q}_{1}\right\} ,\quad N_{2}=\left\{ Q_{2},\bar{Q}%
_{2}\right\} ,  \label{eq:19}
\end{equation}%
and the action of the number operators on the fermionic generator is defined
as%
\begin{eqnarray}
\left[ N_{1},Q_{1}\right]  &=&\left[ N_{1},\bar{Q}_{1}\right] =0,\left[
N_{1},Q_{2}\right] =Q_{2},\left[ N_{1},\bar{Q}_{2}\right] =-\bar{Q}_{2}, 
\notag \\
\left[ N_{2},Q_{2}\right]  &=&\left[ N_{2},\bar{Q}_{2}\right] =0,\left[
N_{2},Q_{1}\right] =Q_{1},\left[ N_{2},\bar{Q}_{1}\right] =-\bar{Q}_{1}.
\label{eq:20}
\end{eqnarray}%
The matrix differential operators act on two-component spinors with
components%
\begin{equation}
P_{N,M}=\left\langle (x^{0},x^{1},\ldots x^{N},),\quad x^{0}\sigma
^{+},x^{1}\sigma ^{+},\ldots x^{M}\sigma ^{+}\right\rangle .  \label{eq:21}
\end{equation}

The linear and bilinear combinations of the operators given in (\ref{eq:17b}%
) Lie algebra is QES, when the space is defined as $P_{n+1,n}$\cite{15}. Let
us consider the following linear combinations of the fermionic generators:%
\begin{equation}
L=2\mu N_{1}+(1+2\mu )N_{2}+\kappa ^{2}(Q_{1}+\bar{Q}_{2})+Q_{2}+\bar{Q}_{1}.
\label{eq:22}
\end{equation}%
Then the eigenvalue problem can be expressed as%
\begin{equation}
L\varphi (x)=\lambda \varphi (x),  \label{eq:23}
\end{equation}%
where%
\begin{equation}
\varphi (x)=\left[ 
\begin{array}{l}
\varphi _{1}(x) \\ 
\varphi _{2}(x)%
\end{array}%
\right]  \label{eq:24}
\end{equation}%
is a two-component spinor.

The algebraic structure of $osp(2,2)$ has been studied previously \cite%
{11,15}. It is obvious that the fermionic generators of $osp(2,2)$ with
number operators $N_{1}$ and $N_{2}$ form a subalgebra of $osp(2,2)$.
Inserting the differential realization (\ref{eq:17b}) into (\ref{eq:22}) and
comparing the equations (\ref{eq:23}) with (\ref{eq:16a}) and (\ref{eq:16b}%
), we can show that the equations are identical when the following hold: 
\begin{subequations}
\begin{eqnarray}
\lambda &=&\frac{1}{2}\left( 1+j+2\mu +2k(1+4\mu )\right) ,  \label{eq:25a}
\\
\epsilon &=&k-\frac{j}{2}-\frac{1}{2}-\kappa ^{2}.  \label{eq:25b}
\end{eqnarray}

The restricted value of $\epsilon $ are the same as those of well-known
results\cite{4}. When the generators act on the space $P_{n+1,n}$ we can
obtain the two recurrence relations 
\end{subequations}
\begin{subequations}
\begin{eqnarray}
nv_{n}+(k-n+\mu -\frac{1}{2}-\frac{j}{2})\omega _{n}+\kappa ^{2}v_{n+1} &=&0,
\label{eq:26a} \\
(2k-n)\omega _{n}+(k-n-\mu -\frac{1}{2}-\frac{j}{2})v_{n+1}+\kappa
^{2}\omega _{n+1} &=&0,  \label{eq:26b}
\end{eqnarray}%
where $\omega _{n}$ and $v_{n}$ are degree $n$ polynomials in $x$ and
constitute a basis for the representation space of the generators of the
algebra. They are related to the eigenfunctions as 
\end{subequations}
\begin{equation}
(P_{n+1},P_{n})=(\varphi _{1},\varphi _{2})=(v_{n+1},\omega _{n}).
\label{eq:27}
\end{equation}

It is necessary that the determinant of these sets be equal to zero, giving
the compatibility conditions that establish the locations of the Juddian
points on the energy baselines. The recurrence relation implies that the
wavefunction is itself the generating function of the energy polynomials.
The eigenvalues are then given simply by the roots of such polynomials. If $%
\kappa _{j}$ is a root of the polynomial $(v_{n+1},\omega _{n})$, the wave
function is truncated at $n=2k$ and belongs to the spectrum of the
Hamiltonian $L$.

\section{Results}

The recurrence relations in (\ref{eq:26a}) and (\ref{eq:26b}), depending on
the choice of the parameters $\mu $ and $j$, describe a number of physical
systems \cite{8}. These recurrence relations can be given in the matrix form:

$\left[ 
\begin{array}{cccccc}
k+\mu -\frac{1+j}{2} & \kappa ^{2} & 0 & 0 & 0 & \cdot \\ 
2k & k-\mu -\frac{1+j}{2} & \kappa ^{2} & 0 & 0 & \cdot \\ 
0 & 1 & k+\mu -\frac{3+j}{2} & \kappa ^{2} & 0 & \cdot \\ 
0 & 0 & 2k-1 & k-\mu -\frac{3+j}{2} & \kappa ^{2} & \cdot \\ 
0 & 0 & 0 & 2 & \cdot & \cdot \\ 
\cdot & \cdot & \cdot & \cdot & \cdot & \cdot%
\end{array}%
\right] \left[ 
\begin{array}{c}
\omega _{0} \\ 
v_{1} \\ 
\omega _{1} \\ 
v_{2} \\ 
\omega _{2} \\ 
\vdots%
\end{array}%
\right] =0.$

The matrix Hamiltonian is identical with the results obtained by Loorits %
\cite{8} when we replace $j$ by $-j-1$ and set $2\mu =G$. If $\mu =0$ and $j$
is replaced by $-j-1,$ then our results are identical to Judd's results \cite%
{4}. The physical systems described by the recurrence relations in (\ref%
{eq:26a}) and (\ref{eq:26b}) after replacing $j$ by $-j-1$, are summarized
as follows.

\begin{quote}
a) $\mu =0$ and $j=0$: \textit{displaced harmonic oscillator.}

b) $\mu =0$ and $j$ is half integer: \textit{linear }$E\otimes \varepsilon $ 
\textit{JT} \textit{system.}

c) $\mu =0$ and $j$ is integer: \textit{linear} $\Gamma _{8}\otimes \tau
_{2} $ \textit{and} $\Gamma _{8}\otimes (\varepsilon +\tau _{2})$ \textit{%
systems.}

d) $\mu \neq 0$ and $j=0$: \textit{dimer.}

e) $\mu \neq 0$ and $j$ is half integer: \textit{linear} $E\otimes
\varepsilon $ \textit{system in an external field.}
\end{quote}
Thus we see that the method described in this article shows that quasi-exact
solutions of a class of JT problems can be obtained as the roots of the
determinant of a matrix of order $2k$. The hypothetical case a) corresponds
to a displaced one-dimensional harmonic oscillator for which the energy
equation (\ref{eq:11}) holds for all $\kappa $:%
\begin{equation}
E=\left( 2k+\frac{3}{2}\right) -\kappa ^{2}.
\end{equation}%
In the cases b) and c), which correspond to three octahedral JT systems,
have common features. The form of all the determinants is the same for all
three systems, provided that in the $\Gamma _{8}\otimes (\varepsilon +\tau
_{2})$ system, the two modes $\varepsilon $ and $\tau _{2}$ are equally
coupled to the electronic state $\Gamma _{8}$. The case $j=0,$ as the
limiting case in the $E\otimes \varepsilon $ JT system, indicates
relationship between $E\otimes \varepsilon $ JT and the Rabi system. It is
known that the Rabi system and the $E\otimes \beta $ JT system occur in the
resonant excitation of double molecules and dimers. The $E\otimes \beta $ JT
system possesses $sl_{2}(R)$ symmetries\cite{16}, which is a subgroup of $%
osp(2,2)$.

Finally, in the presence of an external field, the energies of the
generalized $E\otimes \varepsilon $ JT system can be obtained by solving the
recurrence relations (\ref{eq:26a}) and (\ref{eq:26b}), without any
restriction. These recurrence relations form polynomials in $\kappa ,$ and
the first few of them are given by,%
\begin{eqnarray}
P_{1}(\kappa ) &=&\eta ,  \notag \\
P_{2}(\kappa ) &=&8\eta \kappa ^{2}-(\rho +1)(\eta ^{2}-1), \\
P_{3}(\kappa ) &=&128\eta \kappa ^{2}-8(\eta (3\eta (\rho +1)-4)-4(\rho
+1))\kappa ^{2}+\eta \rho (\rho +2)(\eta ^{2}-4),  \notag
\end{eqnarray}%
for the values $k=0,1/2$ and $1$,respectively. The roots of these
polynomials belong to the spectrum of the Hamiltonian (\ref{eq:1}). In the
last equation, we have defined $j=-(\eta +\rho +2)/2$ and $\mu =(\eta -\rho
)/4.$ Any exact solutions that may be found (even if they are quasi-exact)
would be useful in testing and improving the approximate and numerical
results.

\section{Conclusion}

It is well known that exact solutions of the generalized $E\otimes
\varepsilon $ have a direct practical importance. We have presented a
quasi-exact solution of the generalized $E\otimes \varepsilon $ JT system.
The present work gives a unified treatment of some earlier works. The method
given here can be extended to other JT and multi-dimensional atomic systems.
A further interesting implication of the method is the existence of the
relation between the $E\otimes \varepsilon $ JT system and the two-photon
Rabi Hamiltonian.

The success of our analysis leads us the connection between isolated exact
and quasi-exact solutions. We have shown that the $E\otimes \varepsilon $ JT
system becomes quasi-exactly solvable with an underlying $osp(2,2)$
symmetry. The method given in this paper constitutes the most general Lie
algebraic analysis of the JT interaction for $E\otimes \varepsilon $
systems. We conclude that it is possible to study isolated, exactly solvable
JT and other optical systems in the framework of quasi-exact solvability. In
this way, isolated exact solutions of the JT Hamiltonians are put in the
correspondence with solutions of the QES Hamiltonians. Finally, we mention a
few things about further studies of the JT Hamiltonians. Instead of the
tedious transformations given in \S 1, the fermion-boson realization of the $%
osp(2,2)$ algebra can be constructed and then the $E\otimes \varepsilon $ JT
Hamiltonian can be expressed in terms of the generators of the corresponding
algebra. This construction may lead a simple calculation.


\begin{thebibliography}{99}
\bibitem{1} H. A. Jahn and E. Teller \textit{Proc.R. Soc. London,Ser.} 
\textrm{A}\textbf{161} (1937), 220.

\bibitem{2} H. C. Longuet-Higgins, U. Oepic, M. H. L. Pryce and R. A. Sack 
\textit{Proc.R. Soc. London.} \textrm{A}\textbf{244}, (1958), 1.

\bibitem{3} M. Szopa and A. Ceulemans \textit{J. of Phys. A} \textbf{30}
(1997), 1295.

\bibitem{4} B. R. Judd \textit{J. of Phys. C} \textbf{12} (1979), 1685.

\bibitem{5} H. G. Reik, M. E. St\"{u}lze and M. Doucha \textit{J. of Phys. A}
\textbf{20} (1987), 6327.

\bibitem{6} N. Klenner \textit{J. of Phys. A~}\textbf{19} (1986), 3823.

\bibitem{7} W. Moffitt and W. Thorson \textit{Phys.Rev}. \textbf{168}
(1968), 362.

\bibitem{8} V. Loorits \textit{J. of Phys. C} \textbf{16} (1983), L711.

\bibitem{9} A. V. Turbiner and A. G. Ushveridze \textit{Phys. Lett. A} 
\textbf{126} (1987), 181.

\bibitem{10} A. V. Turbiner \textit{Commun. Math. Phys.} \textbf{118}
(1988), 467.

\bibitem{11} A. V. Turbiner \textrm{``Lie algebras and linear Operators with
invariant subspace'',} \textit{Lie algebras, Chomologies and New Findings in
Quantum Mechanics, Contemporary Mathematics}, AMS, ed. N.Kamran and P.Olver
(1993), pp. 263.

\bibitem{12} V. Brihaye, and P. Kosinski \textit{J.Math.Phys.} \textbf{35}
(1994), 3089.

\bibitem{13} M. A. Shifman \textit{Int. J. Mod. Phys.} \textrm{A}\textbf{4}
(1989), 2897.\qquad

\bibitem{14} A. Gonzales-Lopez, N. Kamran and P. J. Olver \textit{Comm.Math.
Phys.} \textbf{153} (1993), 117.

\bibitem{15} A. V. Turbiner \textrm{``Quasi-Exactly-Solvable Differential
Equations''}, Vol. 3, Chapter 12, (CRC Press 1995), ed. N. Ibragimov.

\bibitem{16} R. Ko\c{c}, M. Koca and H. T\"{u}t\"{u}nc\"{u}ler \textit{J. of
Phys. A} \textbf{35}, (2002), 9425.
\end{thebibliography}
\end{document}